\documentstyle[11pt] {article}
\newlength{\mytopmargin}
\newlength{\myleftmargin}
\begin{document}
\title{
Integration Formulas and Exact Calculations\\ in the Calogero-Sutherland Model}
\author{P.J. Forrester\thanks{e-mail: matpjf@maths.mu.oz.au}
\\
Department of Mathematics \\
University of Melbourne \\
Parkville, Victoria  3052 \\
Australia}
\date{}
\maketitle
\begin{abstract}
Some integration formulas which either occur or are implicit in Ha's recent
exact calculation of some correlations in the Calogero-Sutherland model
are discussed. These integration formulas include the calculation of the
inner product $<0|\rho (0)|\kappa>$ between the density operator acting on an
excited state and the ground state, and a generalization of the Selberg
integral due to Dotsenko and Fateev.
\end{abstract}

\section{Introduction}

The Calogero-Sutherland model refers to quantum particles on a line interacting
via
the $1/r^2$ pair potential. This model has recently been identified with
one-dimensional non-interacting anyons [1-3] and the excitations have been
shown to exhibit fractional statistics [2,4,5]. Furthermore the model exhibits
some remarkable solvability properties which illustrate the latter: the
ground state dynamical density-density correlation and retarded single
particle Green's function can be calculated exactly at all rational couplings
[4-10]. The calculation of these quantities, which is due in its full
generality to Ha [4], uses the theory of Jack symmetric polynomials
introduced by the present author [6-9] to calculate the static correlations
and retarded single particle Green's function at integer couplings.

In this paper we will discuss some integration formulas which either occur or
are implicit in Ha's [4] calculation (all references to Ha below refer to [4]).

\setcounter{equation}{0}
\renewcommand{\theequation}{1.\arabic{equation}}

\section{Calculation of the ground state dynamical density-density correlation
$D(x,t)$ in the finite system}

\setcounter{equation}{0}
\renewcommand{\theequation}{2.\arabic{equation}}
\subsection{Eigenfunctions of $H$}

In periodic boundary conditions, the $1/r^2$ quantum many body Hamiltonian is
given by
\begin{equation}
H:=-\sum_{j=1}^N \; \frac{\partial^{2}}{\partial x^{2}_{j}} +
2\lambda(\lambda-1) \left(\frac{\pi}{L}\right)^{2} \sum_{1 \leq j<k \leq N}
\frac{1}{\sin^{2}
\pi (x_k-x_j) /L}
\end{equation}
The corresponding ground state wave function is [11]
\begin{equation}
| \; 0 > := \psi_0 (x_1, \ldots ,x_N) := \prod\limits_{1 \leq j<k \leq N}
\Big( \sin \pi (x_k-x_j)/L \Big) ^\lambda \; ,
\end{equation}
and the excited states are labelled by partitions
\begin{equation}
\kappa := (\kappa_1, \dots, \kappa_N), \hspace{.3cm}{\rm where}\hspace{.3cm}
\kappa_1 \ge \dots \ge \kappa_N, \hspace{.3cm}{\rm and}\hspace{.3cm} \kappa_j
\in Z \:
(j=1,\dots,N)
\end{equation}
with a unique excited state for each partition [11]. The exact excited states
are, up to normalization, [8]
\begin{equation}
|\;\kappa> := \psi_0 (x_1, \ldots ,x_N)\times \left \{ \begin{array}{ll}
s_\kappa^\lambda(e^{2 \pi i x_1/L},\dots,e^{2 \pi i x_N/L}),&\kappa_N \ge 0 \\
\prod_{l=1}^N e^{2 \pi i \kappa_N x_l /L} s_{\kappa - \kappa_N}^\lambda
(e^{2 \pi i x_1/L},\dots,e^{2 \pi i x_N/L}),&\kappa_N < 0 \end{array} \right.
\end{equation}
where $s_\kappa^\lambda$ is a Jack symmetric polynomial with normalization
chosen so that the coefficient of the monomial symmetric function
\begin{equation}
\sum_{{\rm symmetric} \atop {\rm combination}}
e^{2\pi i x_1\kappa_1/L} e^{2 \pi i x_2\kappa_2/L} \dots e^{2\pi
ix_N\kappa_N/L}
\end{equation}
in its power series expansion is unity [12]. (In [8] we used the notation
$C_\kappa^{(1/\lambda)}$ of Kaneko [13] for the Jack polynomial, which has a
different
normalization to $s_\kappa^\lambda$ ; for our purposes below the latter is more
convenient.)  Also
\begin{equation}
\kappa - \kappa_N := (\kappa_1 - \kappa_N, \kappa_2 - \kappa_N, \dots,
\kappa_N-
\kappa_N)
\end{equation}

\subsection{Eigenfunction expansion of $D(x,t)$}

To calculate the ground state dynamical density-density correlation $D(x,t)$
Ha introduces the eigenstates $|\kappa>$ as a complete set:
\begin{eqnarray}
D(x,t) &:=& <0|\rho (x,t) \rho (0,0)|0> - \rho^2\nonumber \\
& = & \sum_{\kappa,\kappa \not= 0} {<0|\rho (x)|\kappa><\kappa|\rho(0)|0> \over
 <\kappa|\kappa><0 | 0>}
e^{-i(E_\kappa - E_0)t/\hbar} \nonumber \\
&=&\sum_{\kappa,\kappa \not= 0}{|<0|\rho (0) |\kappa>|^2 \over
<\kappa |\kappa><0|0>}
e^{2 \pi i|\kappa|x/L-i(E_\kappa - E_0)t/\hbar}
\end{eqnarray}
where
$$
|\kappa| := \sum_{j=1}^N \kappa_j
\eqno (2.8a)
$$
$$
E_\kappa - E_0 = \left ( 2 \pi \over L \right )^2 \sum_{j=1}^N
( \kappa_j^2 + \lambda \kappa_j (N+1 -2j) )
\eqno (2.8b)
$$
and
$$
\rho(x,t) := e^{-iHt/\hbar}\rho (x) e^{iHt/\hbar}
\eqno (2.8c)
$$
with
$$
\rho(x) = \sum_{j=1}^N \delta(x-x_j)
\eqno (2.8d)
$$
Ha then quotes the value of $<\kappa |\kappa>$, for $\kappa_N \ge 0$, from
Macdonald [14]. With our choice of normalization for the Jack symmetric
polynomial, from Kadell [12], for $\kappa_N \ge 0$
\setcounter{equation}{8}
\begin{equation}
<\kappa |\kappa> = L^N N! {f_N^\lambda(\kappa) \over {\bar
f}_N^\lambda(\kappa)}
\end{equation}
where
$$
f_N^\lambda(\kappa) = \prod_{1 \le i < j \le N}
((j-i)\lambda + \kappa_i - \kappa_j)_\lambda
\eqno (2.10a)
$$
and
$$
{\bar f}_N^\lambda(\kappa) = \prod_{1 \le i < j \le N} (1 - \lambda + (j -
i)\lambda
+\kappa_i - \kappa_j)_\lambda
\eqno (2.10b)
$$
with
$$
(a)_\lambda := {\Gamma (a+\lambda) \over \Gamma (a) }
\eqno (2.10c)
$$

The cases $\kappa_N < 0$ are not mentioned by Ha. In fact {\it provided} the
choice
of normalization implied by $s_\kappa^\lambda$ is made, the following
result is valid.

\vspace{.3cm}
\noindent
\underline{ Proposition 1}

For $\kappa_N < 0$
$$
<\kappa |\kappa> = <\hat{\kappa} |\hat{\kappa}>
\eqno (2.11a)
$$
where
$$
\hat{\kappa} = (-\kappa_N, -\kappa_{N-1},\dots,-\kappa_1)
\eqno (2.11b)
$$
\underline{ Proof}

By definition
$$
<\kappa |\kappa> := \left ( \prod_{l=1}^N \int_0^L dx_l \, \right )
|s_\kappa^\lambda(e^{2 \pi i x_1/L},\dots,e^{2 \pi i x_N/L})|^2
$$
$$
\times
\prod_{1 \le j < k \le N} |e^{2 \pi i x_j/L} - e^{2 \pi i x_k/L}|^{2\lambda}
\eqno (2.12)
$$
But [12]
$$
s_\kappa^\lambda(e^{2 \pi i x_1/L},\dots,e^{2 \pi i x_N/L})
= s_{\hat{\kappa}}^\lambda(e^{-2 \pi i x_1/L},\dots,e^{-2 \pi i x_N/L})
\eqno (2.13)
$$
where the convention, in the case $\kappa_1>0$,
$$
s_{\hat{\kappa}}^\lambda(t_1,\dots,t_N) := \prod_{l=1}^N
t_l^{-\kappa_1}s_{\hat{\kappa}
+\kappa_1}^\lambda(t_1,\dots,t_N)
\eqno (2.14)
$$
is to be adopted
(the formula (2.13) is special to the particular normalization of
$s_\kappa^\lambda$).
The result (2.11a) follows immediately.

\vspace{.3cm}
\noindent
Remark: For $\kappa_N < 0$, from (2.4), it is generally true that
$$
<\kappa|\kappa> = <\kappa-\kappa_N|\kappa-\kappa_N>
$$
Proposition 1 is useful in the cases when both $\kappa_N < 0$ and $\kappa_1 \le
0$.
In such cases all parts of $\hat{\kappa}$ are non-negative so there is an
equality between the norms of states $|\hat{\kappa}>$ labelled by partitions
with non-negative parts, and states $|\kappa>$ labelled by partitions with
non-positive parts.

\vspace{.3cm}

To calculate the inner product $<0|\rho (0)|\kappa>$, Ha uses an expansion of
$\rho (x)$ in terms of Jack symmetric polynomials , and a fundamental
integration formula for Jack symmetric polynomials
(see e.g. [12]). Again only the cases with
$\kappa_N \ge 0$ were considered. Let us give an alternative derivation of the
evaluation of the inner product, applicable for all $|\kappa>$.
\pagebreak

\vspace{.3cm}
\noindent
\underline {Proposition 2}

With $\kappa = -c+\mu$, $(\mu \not= c)$ $\mu_N \ge 0$, $c \in Z_{\ge 0}$
\begin{eqnarray}
<0|\rho (0)|\kappa> &:=&
N L^{N-1}\left ( \prod_{l=1}^N \int_0^1 d\theta_l \,
e^{-2 \pi i \theta_l c}|1 - e^{2 \pi i \theta_l}|^{2\lambda}\right )
s_\mu^\lambda(1,e^{2 \pi i \theta_2},\dots,e^{2 \pi i \theta_N}) \nonumber \\
& & \times \prod_{2 \le j < k \le N} |e^{2 \pi i \theta_k} - e^{2 \pi i
\theta_j}|^{2\lambda}
\nonumber \\
& = & N L^{N-1}{(-1)^{Nc}  (Nc - |\mu|)}N! f_N^\lambda(\mu)\prod_{j=1}^N
\Gamma((N-j)\lambda + 1) \nonumber \\
& & \times \lim_{\epsilon \rightarrow 0}{1 \over \epsilon}
\prod_{j=1}^N { 1 \over \Gamma (-c-\epsilon +1 +(N-j)\lambda + \mu_j)
\Gamma (c+\epsilon +1 +(j-1)\lambda - \mu_j)} \nonumber
\end{eqnarray}
$$
\eqno (2.15)
$$

\noindent
\underline{ Proof}

We start with the fundamental integration formula for Jack polynomials, written
in the form of ref.[12] (with extension to real $a,b,\lambda$ following [15])
$$
\left ( \prod_{l=1}^N \int_{-1/2}^{1/2} d\theta_l \,
e^{ \pi i \theta_l (a-b)}|1 - e^{2 \pi i \theta_l}|^{(a+b)}\right )
s_\mu^\lambda(e^{2 \pi i \theta_1},\dots,e^{2 \pi i \theta_N})
$$
$$
\times \prod_{1 \le j < k \le N} |e^{2 \pi i \theta_k} - e^{2 \pi i \theta_j}|
^{2 \lambda}
$$
$$
= N! f_N^\lambda(\mu)(-1)^\mu \prod_{j=1}^N {\Gamma (a + b + 1 + (N-j)\lambda)
\over \Gamma (a+1 + (N-j)\lambda + \mu_j)\Gamma (b+1 + (j-1)\lambda - \mu_j)}
\eqno (2.16)
$$
We make the substitutions $a=-b$, then $b=c+\epsilon$, where $c \in Z_{\ge 0}$,
and next take the limit $\epsilon \rightarrow 0$. On the l.h.s. we expand the
exponentials involving $\epsilon$:
$$
\prod_{l=1}^N e^{-2 \pi i \theta_l \epsilon}
= 1 - 2 \pi i \epsilon \sum_{l=1}^N \theta_l + O(\epsilon^2)
$$
and thus obtain
$$
<0|\kappa> - 2 \pi i \epsilon \left ( \prod_{l=1}^N \int_{-1/2}^{1/2} d\theta_l
e^{-2 \pi i \theta_l c} \right ) \left ( \sum_{l=1}^N \theta_l \right )
s_\mu^\lambda(e^{2 \pi i \theta_1},\dots,e^{2 \pi i \theta_N})
$$
$$
\times \prod_{1 \le j < k \le N} |e^{2 \pi i \theta_k} - e^{2 \pi i \theta_j}|
^{2 \lambda} + O(\epsilon^2)
\eqno (2.17)
$$

 Since we are assuming $\kappa \not= 0$,
$$
<0|\kappa>=0,
\eqno (2.18)
$$
by orthogonality of the ground state with the excited states. In the
$O(\epsilon)$
term of (2.17), we can use the symmetry of the integrand to replace
$\sum_{l=1}^N \theta_l$
by $N\theta_1$, and then use the subsequent periodicity of the integrand in
$\theta_2, \dots,\theta_N$ to change variables $\theta_j \mapsto \theta_j
+\theta_1$
$(j=2, \dots, N)$. Noting
$$
s_\mu^\lambda(e^{2 \pi i \theta_1},e^{2 \pi i (\theta_2 + \theta_1)},
\dots,e^{2 \pi i (\theta_N + \theta_1)}) =
e^{2 \pi i \theta_1 |\mu|}
s_\mu^\lambda(1,e^{2 \pi i \theta_2},\dots,e^{2 \pi i \theta_N})
$$
the l.h.s. of (2.16) then reads
$$
- 2 \pi i \epsilon \left ( \int_{-1/2}^{1/2} d\theta_1 \theta_1 e^{- 2 \pi i
\theta_1
(Nc - |\mu|)}\right )\left ( \prod_{l=2}^N \int_{-1/2}^{1/2} d\theta_l
e^{-2 \pi i \theta_l c}|1 - e^{2 \pi i \theta_l}|^{2 \lambda} \right )
$$
$$
\times
s_\kappa^\lambda(1,e^{2 \pi i \theta_2},\dots,e^{2 \pi i \theta_N})
\prod_{2 \le j < k \le N} |e^{2 \pi i \theta_k} - e^{2 \pi i \theta_j}|
^{2 \lambda} + O(\epsilon^2)
\eqno (2.19)
$$
After computing the integral over $\theta_1$, and applying the same operations
to the r.h.s. of (2.16), the result (2.15) then follows.

\vspace{.3cm}
Due to the factor $1/(\epsilon \prod_{j=1}^N \Gamma (\epsilon + 1 + (j-1)\gamma
- \mu_j))$ on the r.h.s. of (2.15), we immediately observe from Proposition 1

\vspace{.3cm}
\noindent
\underline{Corollary 1} (Ha)

For $c=0$ and $\lambda = p/q$ ($p$ and $q$ relatively prime positive integers),
$<0|\rho(0)|\kappa>$  is non-zero if and only if
$$
\kappa = (\alpha_1, \dots, \alpha_q,\underbrace{p,\dots,p}_{\beta_1 \,p's},
\dots,\underbrace{1,\dots,1}_{\beta_p \,1's},0,\dots,0)
\eqno (2.20)
$$
where
$$
\alpha_1 \ge \alpha_2 \ge \dots \ge \alpha_q \qquad {\rm and} \qquad
q+\sum_{j=1}^p \beta_j \le N
\eqno (2.21)
$$
Remark: The integers $\alpha_1, \dots, \alpha_q$
are said to label quasi-particle
excitations while the integers $\sum_{j=1}^k \beta_j$ $(k=1,\dots, p)$
are said to label quasi-hole
excitations.

\vspace{.3cm}
In the cases $c>0$, we can deduce from Proposition 1 a similar result.

\vspace{.3cm}
\noindent
\underline{Corollary 2}

\noindent
(i) With $\kappa = -c + \mu$, $c \in Z^+$, $\kappa_N = 0$ and with $\lambda$ as
in
Corollary 1, for $<0|\rho(0)|\kappa>$ to be non-zero $\mu$ must be of the
form
$$
\mu = (c,\dots,c,
\underbrace{c-1,\dots,c-1}_{\beta_1 \,c-1's},\dots,\underbrace{c-p,\dots,c-p}_
{\beta_p \,c-p's}, \alpha_{q-1}, \dots, \alpha_1,0)
\eqno (2.22)
$$

\noindent
(ii) In the cases (2.22),
$$
<0|\rho(0)|\kappa>=<0|\rho(0)|\hat{\kappa}>
\eqno (2.23)
$$
where
$$
\hat{\kappa} = (c,c-\alpha_1,\dots,c-\alpha_{q-1},\underbrace{p,
\dots,p}_{\beta_p \,p's}, \dots, \underbrace{1,\dots,1}_{\beta_1 \, 1's},
0,\dots,0)
\eqno (2.24)
$$

\noindent
\underline{Proof}

\noindent
(i) Using the condition $\kappa_N=0$ and noting that
$$
\lim_{\epsilon \rightarrow 0}{1 \over \epsilon \Gamma (-c-\epsilon +1)} =
(-1)^c \Gamma
(c)
$$
the r.h.s. of (2.15) can be written
$$
N L^{N-1}(-1)^{(N+1)c} \Gamma (c)  (N c - |\mu|) N! f_N^\lambda (\mu )
\prod_{j=1}^N {\Gamma(1+(N-j)\lambda) \over \Gamma(c + 1 + (j-1)\lambda -
\mu_j)}
\prod_{j=1}^{N-1}{1 \over \Gamma (-c+1+(N-j)\lambda + \mu_j)}
\eqno (2.25)
$$
We see immediately from the factor $1/\Gamma (c+1-\mu_1)$ that for (2.25)
to be non-zero we require
$$
\kappa_1 \le c \qquad {\rm and \: thus} \qquad 0 \le \kappa_j \le c \: (j=2,
\dots,N-1)
\eqno (2.26)
$$
Assuming this condition, we see from the second product that (2.25) is non-
zero provided
$$
-c+1+(N-j)\lambda + \mu_j \not= 0,-1,-2,\dots (j=2,\dots,N-1)
\eqno (2.27)
$$
With $\lambda=p/q$ we see that (2.27) holds for $j=N-1, \dots,N-q+1$ so we can
choose
$$
\mu_{N-1} = \alpha_1, \dots, \mu_{N-q+1}= \alpha_{q-1},
\eqno (2.28)
$$
the only constraint being that
$$
0 \le \alpha_1 \le \dots \le \alpha_{q-1} \le b
\eqno (2.29)
$$
In the next case, $j=N-q$, (2.27) reads
$$
-c + 1 + p + \kappa_{N-q} \not= 0,-1, \dots
\eqno (2.30)
$$
Hence we must have
$$
c \ge \kappa_{N-q} \ge c-p
\eqno (2.31)
$$
Assuming this condition and (2.26), (2.27) then holds for all remaining $j$.
The
partitions giving a non-zero value to $<0|\rho(0)|\kappa>$ are therefore given
by (2.20).

\noindent
(ii) To prove this statement we substitute $\mu$ as given by (2.22) in the
r.h.s.
of (2.15), and then take the complex conjugate of the integrand, which we can
do without changing the value of the integral since the latter is real. The
stated result then follows from (2.14) and the fact that $s_\mu^\lambda$ is a
homogeneous function.

\vspace{.3cm}
\noindent
Remark: One consequence of Corollaries 1 and 2 is that the non-zero terms in
(2.7)
have $\kappa$ with all parts non-negative or all parts non-positive.
Furthermore
Corollary 2, together with Proposition 1, (2.8a) and (2.8b) show that in (2.7)
a term having $\kappa$ with all parts non-positive gives the same contribution
as
the term corresponding to $\hat{\kappa}$, which has all parts non-negative,
except
that $\exp (2 \pi i|\kappa|x/L)$ needs to be replaced by $\exp (-2 \pi
i|\kappa|x/L)$.

\section{Normalization of the retarded single particle Green's function}

\setcounter{equation}{0}
\renewcommand{\theequation}{3.\arabic{equation}}
\subsection{An exact result of Ha}

{}From the theory of the above section,
with $\lambda=p/q$, Ha has reported that an exact closed form
expression for $D(x,t)$ can be obtained in the thermodynamic limit. This
exact expression is in the form of a $p+q$ dimensional integral.
 Ha has also announced a similar evaluation of the
retarded single particle Green's function describing hole propogation (for
$\lambda = p$ this formula has also been independently derived by the present
author [9]):
$$
\lim_{N,L \rightarrow \infty \atop N/L = \rho}
<0|\psi^\dagger(x,t)\psi(0,0)|0> \hspace{4cm}
$$
$$
= \rho D e^{-i \pi \lambda \rho x} \left ( \prod_{i=1}^{q-1}\int_0^\infty dx_i
\right )
\left ( \prod_{j=1}^p \int_0^1 dy_j \right ) F(q-1,p,\lambda|\{x_i,y_j\})
$$
$$
\times e^{i (Q_{p,q-1}x-E_{p,q-1}t)}
\eqno (3.1a)
$$
where the momentum $Q$ and the energy $E$ variables are given by
$$
Q_{p,q} := 2 \pi \rho \left ( \sum_{i=1}^q x_i + \sum_{j=1}^p y_j \right )
$$
$$
E_{p,q} := (2 \pi \rho )^2 \left ( \sum_{i=1}^q \epsilon_P(x_j)
+ \sum_{j=1}^p \epsilon_H(y_j) \right )
\eqno (3.1b)
$$
with
$$
\epsilon_P(x) = x(x+\lambda)
$$
$$
\epsilon_H(y) = \lambda y (1-y),
\eqno (3.1c)
$$
the form factor $F$ is given by
$$
F(q,p,\lambda|\{x_i,y_j\}) = \prod_{i=1}^q \prod_{j=1}^p (x_i + \lambda
y_j)^{-2}
{\prod_{i<i'}|x_i - x_i'|^{2 \lambda}\prod_{j<j'}|y_j - y_j'|^{2 /\lambda}
\over \prod_{i=1}^q (\epsilon_P(x_i))^{1-\lambda} \prod_{j=1}^p
(\epsilon_H(y_j))^{1 - 1/\lambda} }
\eqno (3.1d)
$$
and the normalization $D$ is given by
$$
D = {\lambda^{2 p (q-1)} \Gamma (1 + \lambda) \over \lambda^{2p + 1} (q-1)! p!}
A(q-1,p,\lambda)
\eqno (3.1e)
$$
with
$$
A(m,n,\lambda) := {\Gamma^m(\lambda) \Gamma^n(1/\lambda) \over
\prod_{i=1}^m \Gamma^2(p-\lambda(i-1))\prod_{j=1}^n\Gamma^2(1-(j-1)/\lambda)}
\eqno (3.1f)
$$

In this section we focus attention on a corollary of (3.1), following from the
normalization condition
$$
<0|\psi^\dagger(x,t)\psi(0,0)|0> = \rho,
\eqno (3.2)
$$
which is the exact evaluation of the multidimensional integral
$$
 \left ( \prod_{i=1}^{q-1} \int_0^\infty dx_i \right )
\left ( \prod_{j=1}^p \int_0^1 dy_j \right ) F(q-1,p,\lambda|\{x_i,y_j\})
=1/D
\eqno (3.3)
$$

With $q=1$ (3.3) is a special case of the Selberg integral [16] (see e.g. [12]
for a more accessible reference).  However for  $q > 1$ the value of the
integral is not well known. In the next subsection we will show that in this
case (3.3) is a special case of an integral implicit in the work of
Dotsenko and Fateev [17].

\subsection{A generalization of the Selberg integral}

Dotsenko and Fateev [17] consider the multidimensional integral
$$
J_{nm}(\alpha,\beta;\hat{\rho}):=
\left ( \prod_{i=1}^n \int_{{\cal C}_i} dt_i \right )
\left ( \prod_{j=1}^m \int_{{\cal S}_j} d\tau_j \right )
f_{nm}(\{t_i\},\{\tau_j\},\alpha,\beta;\hat{\rho})
\eqno (3.4a)
$$
where
$$
f_{nm}(\{t_i\},\{\tau_j\},\alpha,\beta;\hat{\rho}) \hspace{9cm}
$$
$$
:= \prod_{i=1}^n t_i^{\alpha'}(1 - t_i)^{\beta'} \prod_{j=1}^m
\tau_j^\alpha (1-\tau_j)^\beta { \prod_{i<i'} (t_i-t_{i'})^{2\hat{\rho'}}
\prod_{j<j'} (\tau_j-\tau_{j'})^{2\hat{\rho}} \over
\prod_{i=1}^n \prod_{j=1}^m (\tau_j - t_i)^2 },
\eqno (3.4b)
$$
the multivalued product $\prod_{i<i'}(t_i - t_{i'})^{2 \hat{\rho'}}$ is
defined so that when all the variables $t_i$ are real and ordered $t_1 >
t_2 > \dots >t_n$, the phases of the product are zero (the product
$\prod_{j<j'} (\tau_j - \tau_{j'})^{2/\hat{\rho}} $is defined similarly), and
the contours
${\cal C}_i,{\cal S}_j$ are as in figure 1a. The parameters
$\alpha, \alpha', \beta, \beta'$ and $\hat{\rho}, \hat{\rho'}$ are subject to
the relations
$$
\alpha' = -\hat{\rho'} \alpha \qquad \beta' = - \hat{\rho'} \beta \qquad
\hat{\rho'}
=1/\hat{\rho}
\eqno (3.5)
$$

Interpolating between the formulas (A.15) and (A.16) of [17] we can read off
that
$$
J_{nm}(\alpha,\beta;\rho)=
\prod_{j=0}^{n-1} { \sin \pi ( 2 - 2m + \alpha' + \beta' + \hat{\rho'}
(n - 1 + j)) \over \sin \pi (1 + \alpha' + j \rho' ) }
$$
$$
\times \tilde{J}_{nm}^{(n)}(\alpha, \beta;\hat{\rho})
\eqno (3.6a)
$$
where
$$
 \tilde{J}_{nm}^{(n)}(\alpha, \beta;\hat{\rho})
 :=
\left ( \prod_{i=1}^n \int_{{\cal K}_i} dt_i \right )
\left ( \prod_{j=1}^m \int_{{\cal S}_j} d\tau_j \right )
\tilde{f}_{nm}({t_i},{\tau_j},\alpha,\beta;\hat{\rho})
\eqno (3.6b)
$$
with $\tilde{f}_{nm}$ defined as in (3.4b) except that each factor
$(1-t_i)^{\beta'}$
is to be replaced by $(t_i-1)^{\beta'}$, and the contours ${\cal K}_i$
are as in figure
1b. Furthermore, all the contours on the l.h.s. and r.h.s. of (3.6a) can be
collapsed onto the real axis. Extra "phase factors", due to the products
$$
\prod_{i<i'} (t_i-t_{i'})^{2\hat{\rho'}} \qquad {\rm and} \qquad
\prod_{j<j'} (\tau_j-\tau_{j'})^{2\hat{\rho}}
$$
result, but are the same for both sides of the equation and so cancel. We thus
have
$$
\tilde{I}_{nm}(\alpha, \beta;\hat{\rho})=
\prod_{j=0}^{n-1}
{\sin \pi (1 + \alpha' +j\hat{\rho'}) \over
\sin \pi ( 2 - 2m + \alpha' + \beta' + \hat{\rho'}(n-1 + j))}
I_{nm}(\alpha, \beta;\hat{\rho})
\eqno (3.7a)
$$
where
$$
\tilde{I}_{nm}(\alpha, \beta;\hat{\rho}) :=
\left ( \prod_{i=1}^n \int_1^\infty dt_i \right )
\left ( \prod_{j=1}^m \int_0^1 d\tau_j \right )
|f_{nm}({t_i},{\tau_j},\alpha,\beta;\hat{\rho})|
\eqno (3.7b)
$$
and
$$
I_{nm}(\alpha, \beta;\hat{\rho}) := {\rm P}
\left ( \prod_{i=1}^n \int_0^1 dt_i \right )
\left ( \prod_{j=1}^m \int_0^1 d\tau_j \right )
|f_{nm}({t_i},{\tau_j},\alpha,\beta;\hat{\rho})|
\eqno (3.7c)
$$
where P stands for the principal value integral. The integral (3.7c) is
evaluated by (A.35) of [17] (there is a minor misprint in this equation:
in the final two products the lower terminals should be 0 instead of 1),
so inserting its value in (3.7a) and using the formula
$$
\Gamma (z) \Gamma (1-z) ={ \pi \over \sin \pi z}
$$
we conclude
\setcounter{equation}{7}
\begin{eqnarray}
\tilde{I}_{nm}(\alpha, \beta;\hat{\rho})& =&
m! n! \hat{\rho}^{2 n m} \prod_{l=1}^n {\Gamma(l\hat{\rho'}) \over \Gamma
(\hat{\rho'})} \prod_{j=1}^m {\Gamma(j\hat{\rho}-n) \over \Gamma
(\hat{\rho})} \nonumber \\
& & \times \prod_{l=0}^{n-1} { \Gamma (1+\beta' +l\hat{\rho'})
\Gamma (-1 + 2m -\alpha' - \beta' - (n-1+l)\hat{\rho'}) \over
\Gamma (-\alpha' - l \hat{\rho'}) }\nonumber \\
& & \times \prod_{j=0}^{m-1} { \Gamma (1-n+\alpha +j\hat{\rho})
\Gamma(1 - n + \beta +j\hat{\rho}) \over
\Gamma(2 - n + \alpha + \beta + (m-1+j)\hat{\rho}) }
\end{eqnarray}
This is the sought generalization of the Selberg integral.

It is straightforward to express the l.h.s. of (3.3) in terms of this integral:
$$
 \left ( \prod_{i=1}^{q-1}\int_0^\infty dx_i \right )
\left ( \prod_{j=1}^p \int_0^1 dy_j \right ) F(q-1,p,\lambda|\{x_i,y_j\})
= {1 \over \lambda^{pq-1}} \tilde{I}_{q-1,p}(q/p-1,q/p-1,q/p)
\eqno (3.9)
$$
That the evaluations given by (3.3) and (3.8) agree in (3.9) is demonstrated in
the appendix.

\section{Conjecture for the static two-particle distribution function}
\setcounter{equation}{0}
\renewcommand{\theequation}{4.\arabic{equation}}

As our final result we will combine our knowledge of the integrals (2.15) and
(3.8) with a conjecture of Ha to conjecture an integral formula for the static
two
particle distribution function
$$
\rho^{(2)}(x) := {\rm TL} {<0|\sum_{j \not= k} \delta (x-x_j) \delta (0-x_k)|0>
\over <0|0>},
\eqno (4.1)
$$
for rational coupling, where TL denotes the thermodynamic limit.
The conjecture states that the so called form factor, which for our purpose is
the function $F$ in the integrand of (3.1a), is given by (3.1d) for any two
point
correlation function in which the intermediate states involve only
$q$ quasi-particles and $p$ quasi-holes (recall the remark after Corollary 1).

To apply the conjecture, we note that for a system of $N+2$ particles
\setcounter{equation}{1}
\begin{eqnarray}
\lefteqn{<0|\sum_{j \not= k} \delta (x-x_j) \delta (0-x_k)|0>} \nonumber \\
&=& c_{NL}(\lambda)|\sin \pi x/L|^{2\lambda}
\left (\prod_{l=1}^N \int_0^1 dx_l |1-e^{2 \pi i x_l/L}|^{2\lambda}
|1 - e^{2 \pi i (x_l - x)/L}|^{2\lambda} \right ) \nonumber \\
& & \hspace{3cm} \times \prod_{1 \le j < k \le N}
|e^{2 \pi i x_k/L} - e^{2 \pi i x_j/L}|^{2\lambda} \nonumber \\
&=& c_{NL}(\lambda)|\sin \pi x/L|^{2\lambda}
\sum_{\kappa} {|<\kappa|\hat{\Phi}>|^2 \over <\kappa|\kappa>}
 e^{2 \pi i x |\kappa|/L}
\end{eqnarray}
where
$$
|\hat{\Phi}> := \prod_{l=1}^N|1 - e^{2 \pi i x_l/L}|^{2\lambda}
\prod_{1 \le j < k \le N}
|e^{2 \pi i x_k/L} - e^{2 \pi i x_j/L}|^{\lambda}
\eqno (4.3)
$$

{}From Corollary 1 we see that, with $\lambda=p/q$,$<\kappa|\hat{\Phi}>$ is
non-
zero only if $\kappa$ consists of $2(q-1)$ quasi-particle and $2p$ quasi-hole
labels. For example, if $\lambda = 1/2$, the states giving a non-zero
contribution
to (4.2) are, in the cases $\kappa_N \ge 0$, of the form
$$
\kappa = (q_1,q_2,\underbrace{2,\dots,2}_{p_2 \; 2's},\underbrace{1,\dots,1}_
{p_1 \; 1's},0,\dots,0)
\eqno (4.4a)
$$
where
$$
q_1 \ge q_2  \qquad {\rm and} \qquad p_2+p_1 \le N-2.
\eqno (4.4b)
$$

The conjecture of Ha can now be applied to predict
$$
\rho^{(2)}(x)= \rho^2 C(\lambda) (\rho x)^\Gamma
 \left ( \prod_{i=1}^{2(q-1)}\int_0^\infty dx_i \right )
\left ( \prod_{j=1}^{2p} \int_0^1 dy_j \right )
F(2(q-1),2p,\lambda|\{x_i,y_j\})
\cos (Q_{2p,2(q-1)} x)
\eqno (4.5)
$$

It remains to calculate the normalization $C(\lambda)$. For this purpose we
recall
from our previous work [6,eq.(5.23b)] the small-$x$ result
$$
\lim_{x \rightarrow 0^+} {\rho^{(2)}(x) \over \rho^2 (\rho x)^{2 \lambda}}
=A(\lambda)
\eqno (4.6a)
$$
where
$$
A(\lambda) := (2 \pi  \lambda)^{2 \lambda} { (\lambda !)^3 \over
(2 \lambda)! (3\lambda)! }
\eqno (4.6b)
$$
Furthermore
$$
 \left ( \prod_{i=1}^{2(q-1)}\int_0^\infty dx_i \right )
\left ( \prod_{j=1}^{2p} \int_0^1 dy_j \right )
F(2(q-1),2p,\lambda|\{x_i,y_j\})
\hspace{4cm}
$$
$$
= \lambda^{-4pq + 2p/q} \tilde{I}_{2(q-1),2p}(q/p-1,q/p-1,q/p)
\eqno (4.7)
$$
where $\tilde{I}$ is defined by (3.7b) and evaluated by (3.8). Hence the
normalization is given explicitly by
$$
C(\lambda) = {A(\lambda) \over
 \lambda^{-4pq + 2p/q} \tilde{I}_{2(q-1),2p}(q/p-1,q/p-1,q/p)}
\eqno (4.8)
$$

In the special case $\lambda = p$ the conjecture (4.5) agrees with the known
exact evaluation [7].

\vspace{2cm}
\noindent
{\bf Acknowledgements}

\noindent
Some of this work was done during a visit to the Laboratoire de Physique
Th\'eorique at Orsay organised by B. Jancovici, whom I thank for his efforts.
Financial support was provided by the CNRS and the ARC. Also I thank T. Miwa
for
a critical reading of the manuscript.

\pagebreak

\noindent
{\bf Appendix}
\vspace{.5cm}

\noindent
According to (3.3) and (3.8), (3.9) says that for $\lambda = p/q$,
$$
{\lambda^{p-q+1}(q-1)!p! \prod_{i=1}^{q-1} \Gamma^2 (p - \lambda (i-1))
\prod_{j=1}^p \Gamma^2(1 - (j-1)/\lambda) \over
\lambda^{2 p (q-1)} \Gamma (1+ \lambda) \Gamma^{q-1}(\lambda)
\Gamma^p(1/\lambda)} \hspace{3cm}
$$
$$
= (q-1)! p! \lambda^{-2(q-1)p} \prod_{l=1}^{q-1} {\Gamma (l\lambda)
\over \Gamma (\lambda)} \prod_{j=1}^p {\Gamma (j/\lambda - (q-1))
\over \Gamma (1/\lambda)}
$$
$$
\hspace{3cm} \times
\prod_{l=0}^{q-2} {\Gamma ((l+1)\lambda) \Gamma (1 + p - l\lambda)
\over \Gamma (1 - (l+1)\lambda) }
\prod_{j=0}^{p-1} {\Gamma^2(1 - q + (j+1)/\lambda) \over
\Gamma (1 + (j+1)/\lambda)}
\eqno ({\rm A}1)
$$
Here we will show how to reduce the r.h.s. of this identity to the l.h.s..
First, by replacing $l \rightarrow l+1$ and $j \rightarrow j+1$ in the second
last and last products, the r.h.s reads
$$
(q-1)! p! \lambda^{-2(q-1)p} {\Gamma (1+p) \over \Gamma (1+p-(q-1)\lambda)}
\prod_{l=1}^{q-1} {\Gamma^2 (l\lambda) \Gamma (1+p-l\lambda) \over \Gamma (1-l
\lambda)\Gamma (\lambda) }\prod_{j=1}^p {\Gamma^3(1-q+j/\lambda) \over
\Gamma (1+j/\lambda)\Gamma (1/\lambda)}
\eqno ({\rm A}2)
$$
Now
\setcounter{equation}{2}
\renewcommand{\theequation}{A\arabic{equation}}

\begin{eqnarray}
\prod_{l=1}^{q-1} {\Gamma (1+p-l\lambda) \over \Gamma (1-l
\lambda) } & = & \prod_{l=1}^{q-1} (p-l\lambda)(p-1-l\lambda)\dots (1-l\lambda)
\nonumber \\
& = & \lambda^{p(q-1)} \prod_{j=1}^p {\Gamma (j/\lambda) \over \Gamma(j/\lambda
-
(q - 1))}
\end{eqnarray}
and
$$
\prod_{j=1}^p {1 \over \Gamma (1 + j/\lambda)} = {\lambda^p \over p!}
\prod_{j=1}^p{1 \over \Gamma(j/\lambda)}
\eqno ({\rm A}4)
$$
Substituting (A3) and (A4) in (A2) reproduces the l.h.s. of (A1), as required.

\pagebreak
\noindent
{\bf References}
\begin{description}
\item[][1] L. Brink, T.H. Hansson, S. Konstein and M.A. Vasiliev, Nucl. Phys. B
{\bf 401} (1993) 591.
\item[] [2] F.D.M. Haldane, in {\it Proceedings of the 16th Taniguchi
Symposium}
eds. A. Okiji and N. Kawakami, Springer-Verlag, 1994.
\item[] [3] S.B. Isakov, Int. J. of Mod. Phys. A {\bf 9} (1994) 2563.
\item[] [4] Z.N.C. Ha, "Exact Dynamical Correlation Functions of
 Calogero-Sutherland Model and One-dimensional Fractional Statistics",
submitted Phys. Rev. Lett.
\item[] [5] E.R. Mucciolo, B.S. Shastry, B.D. Simons and B.L. Altshuler, Phys.
Rev.
B {\bf 49} (1994) 15197.
\item[] [6] P.J. Forrester, Nucl. Phys. B {\bf 388} (1992) 671.
\item[] [7] P.J. Forrester, Phys. Lett. A {\bf 179} (1993) 127.
\item[] [8] P.J. Forrester, Nucl. Phys. B {\bf 416} (1994) 377.
\item[] [9] P.J. Forrester, "Exact Calculation of the Ground State
Single-Particle
Green's Function for the $1/r^2$ Quantum Many Body System", submitted J. Math.
Phys.
\item[] [10] D.V. Khveshchenko, "Towards exact bosonization of the
Calogero-Sutherland
model", preprint.
\item[] [11] B. Sutherland, Phys. Rev. A{\bf5}, (1972) 1372.
\item[] [12] K.W.J. Kadell, Compos. Math. {\bf 87}, (1993) 5.
\item[] [13] J. Kaneko, SIAM J. Math. Anal. {\bf 24}, (1993) 1086.
\item[] [14] I.G. Macdonald, {\it Symmetric Functions and Hall Polynomials},
2nd ed.,
Oxford University Press, to appear.
\item[] [15] W.G. Morriss, {\it Constant term identities for finite and affine
root
systems}, Ph.D. thesis,  Univ. of Wisconsin, Madison, 1982.
\item[] [16] A. Selberg, Norsk. Mat. Tidsskr. {\bf 26}, (1994) 71.
\item[] [17] V.S. Dotsenko and V.A. Fateev, Nucl. Phys. B {\bf 25}, (1985) 691.
\end{description}

\end{document}